\documentclass[fleqn,usenatbib,useAMS]{mnras}

\usepackage{amsmath,amsfonts,amssymb,bm,graphicx,graphics,tabularx,natbib,color,multicol,pdflscape}
\usepackage{siunitx,threeparttable,pdflscape}

\usepackage{cleveref}

\begin{document}

\title{\textbf{Mass Discrepancy-Acceleration Relation in Einstein Rings}}

\author[Tian \& Ko]
{Yong Tian$^{1}$\thanks{E-mail:yongtian@astro.ncu.edu.tw}
and Chung-Ming Ko$^{1,2}$\thanks{E-mail:cmko@astro.ncu.edu.tw}
\\
$^{1}$Institute of Astronomy, National Central University, Taoyuan City, Taiwan 32001, Republic of China \\
$^{2}$Department of Physics and Center for Complex Systems, National Central University, Taoyuan City, Taiwan 32001, Republic of China
}

\date{Accepted Received ; in original form 2017 Apr 14}

\maketitle

    \begin{abstract}
    We study the Mass Discrepancy-Acceleration Relation (MDAR) of 57 elliptical galaxies by their Einstein rings from the Sloan Lens ACS Survey (SLACS).
    The mass discrepancy between the lensing mass and the baryonic mass derived from population synthesis is larger when the acceleration of the elliptical galaxy lenses is smaller.
    The MDAR is also related to surface mass density discrepancy.
    At the Einstein ring, these lenses belong to high-surface-mass density galaxies.
    Similarly, we find that the discrepancy between the lensing and stellar surface mass density is small.
    It is consistent with the recent discovery of dynamical surface mass density discrepancy in disk galaxies where the discrepancy is smaller when surface density is larger.
    We also find relativistic modified Newtonian dynamics (MOND) can naturally explain the MDAR and surface mass density discrepancy in 57 Einstein rings.
    Moreover, the lensing mass, the dynamical mass and the stellar mass of these galaxies are consistent with each other in relativistic MOND.
    \end{abstract}

\begin{keywords}
gravitation -- gravitational lensing: strong -- galaxies: elliptical and lenticular, cD-- galaxies: kinematics and dynamics -- dark matter
\end{keywords}

    \section{Introduction}\label{sec:intro}
    The mass discrepancy or the missing mass of galactic systems refers to the excess of its dynamical mass over its baryonic mass.
    The mass discrepancy has a tight relation with the observed acceleration $g$ \citep{Sanders90,McGaugh99,McGaugh04} and with the baryonic Newtonian gravitation $g_{\rm N}$ \citep{McGaugh04,TC09,FM12}.
    The discrepancy is larger when the gravitational acceleration of the spiral galaxy is smaller, and the relation is called the mass discrepancy-acceleration relation (MDAR).
    Later, the MDAR is also confirmed by the dynamics of elliptical galaxies \citep{TC16,Janz16}.
    However, mass discrepancy shows no clear relation to other observational quantities such as distance or orbital angular speed (see \cite{McGaugh04} for details).
    The MDAR can be also interpreted as gravitational acceleration discrepancy between $g$ and $g_{\rm N}$.
    Recently, \cite{McGaugh16} found a tight radial acceleration relation (RAR) between $g$ and $g_{\rm N}$ in 153 disk galaxies from Spitzer Photometry and Accurate Rotation Curves (SPARC) database.
    This relation suggested that
    \begin{equation}\label{eq:MOND}
      g/g_{\rm N}=\nu(g_{\rm N}/\mathfrak{a}_0)\,,
    \end{equation}
    where $\mathfrak{a}_0\approx1.2\times10^{-10}$ ms$^{-2}$ and $\nu(y)$ has the asymptotic behavior $\nu(y)\approx 1$ for $y\gg 1$ and $\nu(y)\approx y^{-1/2}$ for $y\ll 1$.
    In modified Newtonian dynamics (MOND), $\nu(y)$ is known as the (inverted) interpolating function.
    For example, a commonly used form, the simple form \citep{FB05},
    \begin{equation}\label{eq:Simple}
      \nu(y)=[1+(1+4y^{-1})^{1/2}]/2\,.
    \end{equation}
    This form will be used for later discussions.

    The surface mass density discrepancy also shows the similar trend as the MDAR because the gravitational acceleration is related to surface mass density as $\Sigma=M/\pi r^2\approx g/\pi G$.
    The discrepancy increases as surface mass density decreases when it is smaller than the characteristic surface mass density $\Sigma_0=\mathfrak{a}_0/\pi\,G$.
    For high surface mass density spiral galaxies ($\Sigma>\Sigma_0$), the mass discrepancy is small.
    For example, recently, \cite{Genzel17} discovered six high redshift spiral galaxies dominated by baryons belongs to high surface mass density galaxies.
    \cite{Milgrom17} explained this in the context of MOND.
    For low surface mass density ones ($\Sigma<\Sigma_0$), the mass discrepancy is large (for review, see, e.g., \cite{SandersMcGaugh,FM12}).
    In fact, the same trend happens in elliptical galaxies, for instances, high surface mass density elliptical galaxies probed by planetary nebulae have small mass discrepancy \citep{Romanowsky03,MS03,TC16}, and for low surface mass density tidal dwarfs the mass discrepancy is large \citep{Gentile07,Milgrom07,Dabringhausen16}.
    Recently, from the surface mass density in the central regions of 135 disk galaxies (S0 to dIrr), \cite{Lelli16} showed that the mass discrepancy increases as surface mass density decreases.
    \cite{Milgrom16} also explained this in the context of MOND.

    The MDAR posed a puzzle in standard $\Lambda$CDM cosmology \citep{WuKroupa15}.
    Several attempts have been made in this direction, both hydrodynamical simulations \citep{WuKroupa15,Ludlow17} and semi-empirical models \citep{DiCintioLelli16, Navarro16,Desmond17}.
    Stochastic galaxy formation should explain four important issues raised by the MDAR and RAR:(1) the characteristic acceleration scale $\mathfrak{a}_0$, (2) the low acceleration relation (the Tully-Fisher relation and the Faber-Jackson relation), (3) the low scatter of the relation, and (4) the lack of correlations between galaxy properties and residuals (see, e.g., Sec 8.2 of Ref. \cite{Lelli17} for the details).

    The mass discrepancy problem appears not only in stellar dynamics of galaxies but also in relativistic phenomena such as gravitational lensing (the light path bending by a massive object predicted by General Relativity, GR).
    For instances, in strong gravitational lensing, the observed angle of deflection of light from a distance source (e.g., a quasar or galaxy) by a gravitational lens (e.g., a galaxy or cluster of galaxies) is larger than the one expected by GR if only the luminous mass from the lens is considered.

    To study relativistic problems in MOND is beyond the modified Poisson equation proposed for non-relativistic dynamics in \cite{BM84}.
    The difficulty is not only the theoretical complications but also the enhanced angle of deflection is not easily satisfied by the usual conformal metric (see, e.g., the discussion in \cite{BS94}).
    In 2004, Bekenstein adopted a disformal metric and proposed the Tensor-Vector-Scaler theory (T$e$V$e\,$S) \citep{Bekenstein04}.
    This is the first covariant relativistic gravitational theory of MOND.
    The angle of deflection has the same formulation in T$e$V$e\,$S as in GR but using MONDian gravitational potential instead (see, e.g., \cite{Chiu06,Tian13} for details).
    For other relativistic MOND theories, such as GEA \citep{Zlosnik07} and BIMOND \citep{Milgrom09}, gravitational lensing result is the same as in \cite{Chiu06} for spherical symmetry case.
    Thus, one may expect the mass discrepancy in relativistic MOND will have the same trend as in non-relativistic MOND.
    MDAR is also expected in gravitational lensing.

    Because strong lenses belong to high surface mass density galaxies (see, e.g., \cite{Sanders14} for details), small mass discrepancy is expected in relativistic MOND.
    When comparing with initial mass function (IMF), MOND can explain this small discrepancy without dark matter \citep{Chen06,Chiu06,Chiu11,Sanders08,Sanders14}.

    It is interesting to study mass discrepancy-acceleration relation in gravitational lensing.
    In Section 2, we describe our data and model.
    In Section 3, we present three results: the MDAR, surface mass discrepancy, and consistency between dynamical and lensing mass in relativistic MOND.

    \section{Data and Model}

        \subsection{Data}

    We select strong lens data from the Sloan Lens ACS (SLACS) database \citep{Auger09}.
    Sloan Digital Sky Survey (SDSS) has observed millions of galaxies.
    When two galaxies are lying close to a line-of-sight with one at a much further distance than the other, it will provide a candidate for strong gravitational lensing, in particular, an Einstein ring if the two galaxies are lying exactly on one line-of-sight.
    The SLACS used the Advanced Camera for Surveys (ACS) of the Hubble Space Telescope photometry to resolve the galaxy lenses.
    Combining with redshift measurements, stellar velocities, and brightness by SDSS, SLACS provided 85 high-quality Einstein rings \citep{Auger09}.

    In this work, we select elliptical galaxy lenses that can be approximated by spherically symmetric mass distribution, with complete photometric data and estimation of stellar mass by population synthesis.
    We also exclude S0 galaxies because of the mass model.
    As a result, we have 57 Einstein rings in our samples, see Table \ref{tab:data}.
    The samples include the size of the Einstein ring $\theta_{\rm Obv}$, the effective radius (or half-light radius) of the lens $R_{\rm eff}$, and the stellar mass (i.e., baryonic mass or luminous mass) ${\cal M}_{\rm b}$ estimated by population synthesis with Salpeter IMF \citep{Auger09}.

        \subsection{Model}

    Assuming the thin-lens approximation, the deflection angle can be written as
    \begin{equation}\label{deflection}
       \alpha(\theta)=\frac{2}{c^2}\int^{\infty}_{-\infty} \nabla_{\bot}\Phi\,{\rm d}s\,,
    \end{equation}
    where $c$ is the speed of light, $s$ is the distance along the light path, $\Phi$ is the gravitational potential, and $\nabla_{\bot}$ is the two-dimensional gradient operator perpendicular to light propagation.
    For the Einstein ring, the lens equation is given by
    \begin{equation}\label{lensEq}
        \theta=\alpha(\theta)\frac{D_{\rm LS}}{D_{\rm S}}\,,
    \end{equation}
    where $D_{\rm L}$, $D_{\rm S}$ and $D_{\rm LS}$ are the angular diameter distances of the lens from the observer, the observer from the source, and the source from the lens, respectively.

    We adopt Hernquist mass model \citep{Hernquist90} for the luminous mass of the elliptical galaxy lenses.
    The distributions of luminous mass or stellar mass and the corresponding Newtonian gravitational acceleration are
    \begin{equation}\label{eq:Hernquist}
      m_{\rm b}(r)=\frac{{\cal M}_{\rm b}r^2}{(r+r_h)^2}\,,
      \quad g_{\rm b}(r)=\frac{G{\cal M}_{\rm b}}{(r+r_h)^2}\,,
    \end{equation}
    with $r_h\approx 0.551 R_{\rm eff}$.

    \section{Results and Discussion}

        \subsection{Mass Discrepancy-Acceleration Relation in Einstein Rings}

        \begin{figure*}
            \includegraphics[width=\columnwidth]{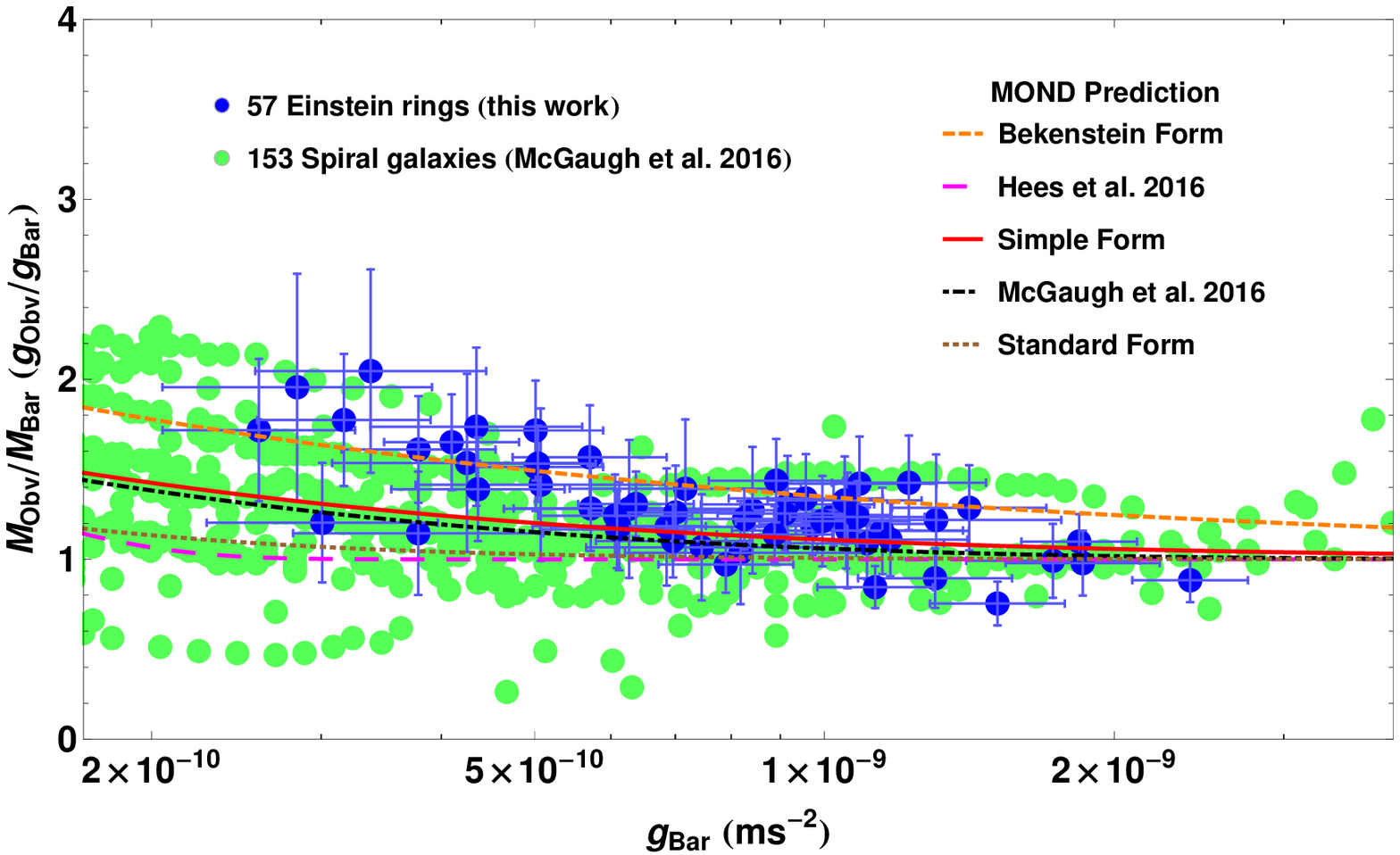}
            \includegraphics[width=\columnwidth]{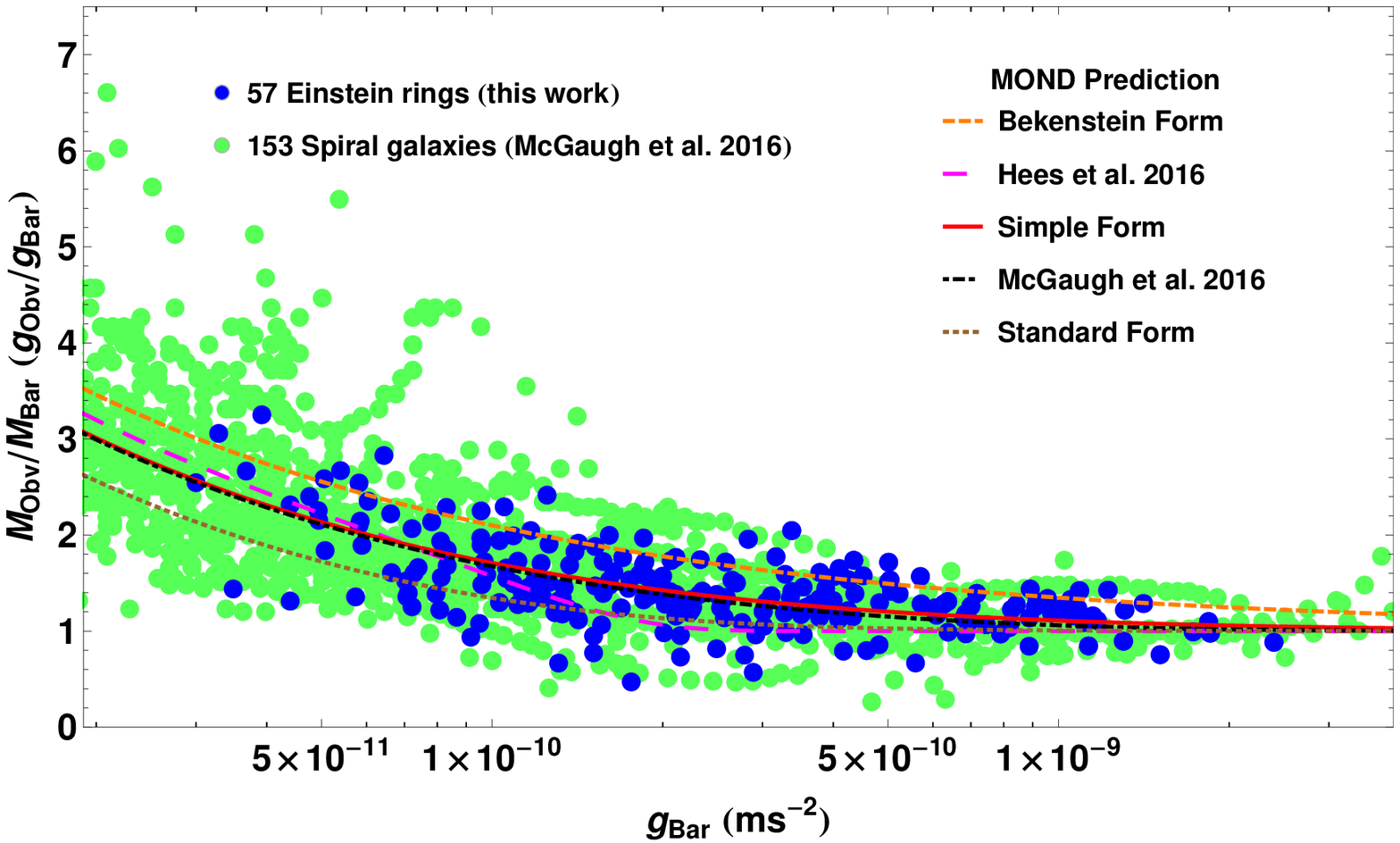}
            \caption{Mass discrepancy-acceleration relation. Blue filled circles are the 57 Einstein rings in this work, and green filled circles are the data of spiral galaxies in
            McGaugh et al. (2016). The horizontal axis is the Newtonian acceleration $g_{\rm Bar}$ (in logarithmic scale) estimated from the baryonic mass $M_{\rm Bar}$ at the effective radius (adopt Hernquist model). For comparison, we plotted the prediction of MOND. The orange dashed, magenta long dashed, red solid, black dot-dashed, and brown dotted lines represent the Bekenstein form, Hees form, simple form, McGaugh form, and standard forms in MOND, respectively. Error bar comes from the error of total baryonic mass estimation. Data and errors are listed in Table \ref{tab:data}. Left panel: Mass discrepancy (or acceleration discrepancy) estimated by $M_{\rm Obv}/M_{\rm Bar}=g_{\rm Obv}/g_{\rm Bar}$ at the effective radius. $M_{\rm Obv}$ is the total mass including stellar mass ($M_{\rm Bar}$) (i.e., baryonic mass) estimated by population synthesis with Salpeter IMF (Auger et al. 2009) and an isothermal sphere dark matter component (see text for details). Right panel: Mass discrepancy-acceleration relation at different radius (adopting Hernquist model) from Einstein ring to 4 effective radii (285 data points).}\label{fig:MassDis}
        \end{figure*}

    To examine the MDAR in our samples, we use two ways to estimate the ratio of the gravitational acceleration from observation to that inferred from luminous mass:
    (1) to compare the angles of deflection, and
    (2) to compare the estimated values of gravitational acceleration at the effective radius.

    Eq. \ref{deflection} indicates that deflection angle represents an average of the gravitational acceleration over the line-of-sight.
    For the Einstein ring, $\alpha_{\rm Obv}/\alpha_{\rm Bar}=\theta_{\rm Obv}/\theta_{\rm Bar}$.
    Here $\theta_{\rm Obv}$ stands for the observed radius of the Einstein ring and $\theta_{\rm Bar}$ for the expected ring radius produced by the baryonic (luminous) mass only, Eq. \ref{eq:Hernquist}.
    Table \ref{tab:data} listed the ratio of $g_{\rm Bar}=g_{\rm b}(R_{\rm eff})$ in Eq. \ref{eq:Hernquist}  (i.e., the Newtonian gravitational acceleration at $R_{\rm eff}$ by the luminous mass only).
    The effective radius $R_{\rm eff}$ of our samples is also listed in Table \ref{tab:data}.
    This result is very close to the second method and the MDAR holds.

    To estimate the mass within the ring radius, we add a dark matter component $m_{\rm dm}(r)=2\sigma_v^2 r/G$ (singular isothermal sphere profile) to the luminous matter (Eq. \ref{eq:Hernquist}).
    $\sigma_v^2$ can be obtained from the observed size of the Einstein ring.
    We plot the ratio $M_{\rm Obv}/M_{\rm Bar}=g_{\rm Obv}/g_{\rm Bar}$ against $g_{\rm Bar}$ in the left panel of Fig. \ref{fig:MassDis}.
    Here, $M_{\rm Bar}=m_{\rm b}(R_{\rm eff})$, $M_{\rm Obv}=M_{\rm Bar}+m_{\rm dm}(R_{\rm eff})$ and $g_{\rm Obv}=GM_{\rm Obv}/R_{\rm eff}^2$.

    As shown in Fig. \ref{fig:MassDis}, the mass discrepancy represented by $M_{\rm Obv}/M_{\rm Bar}$ (left panel) increases as the Newtonian acceleration $g_{\rm Bar}$ decreases.
    If we choose mass or acceleration in radius other than the effective radius in the second method, the MDAR still holds (see the right panel of Fig. \ref{fig:MassDis}).
    Our result is consistent with the result from spiral galaxies reported by \cite{McGaugh16} (see Fig. \ref{fig:MassDis}).
    Our analysis shows the MDAR holds in the relativistic phenomenon, strong gravitational lensing.

    For comparison, in Fig. \ref{fig:MassDis} we plot different (inverted) interpolating functions in MOND
    $\nu(g_{\rm Bar}/\mathfrak{a}_0)=g_{\rm Obv}/g_{\rm Bar}=M_{\rm Obv}/M_{\rm Bar}$ as a function of $g_{\rm Bar}$.
    The orange dashed is Bekenstein form ($\nu(y)=1+y^{1/2}$), the magenta long dashed is Hees form ($\nu(y)=(1-e^{-y^{2}})^{-1/4}+3/4e^{-y^{2}}$) \citep{Hees16}, the red solid line is simple form (see Eq. (\ref{eq:Simple})), the black dot-dashed is McGaugh form ($\nu(y)=e^{\sqrt{y}}/(e^{\sqrt{y}}-1)$) \citep{McGaugh16} and the dotted-line is standard form ($\nu(y)=[(1+4y^{-2})^{-1/2}/2]^{-1/2}$).
    One can see that MOND is consistent to the MDAR of Einstein rings and spiral galaxies.

\begin{figure*}
            \includegraphics[width=\columnwidth]{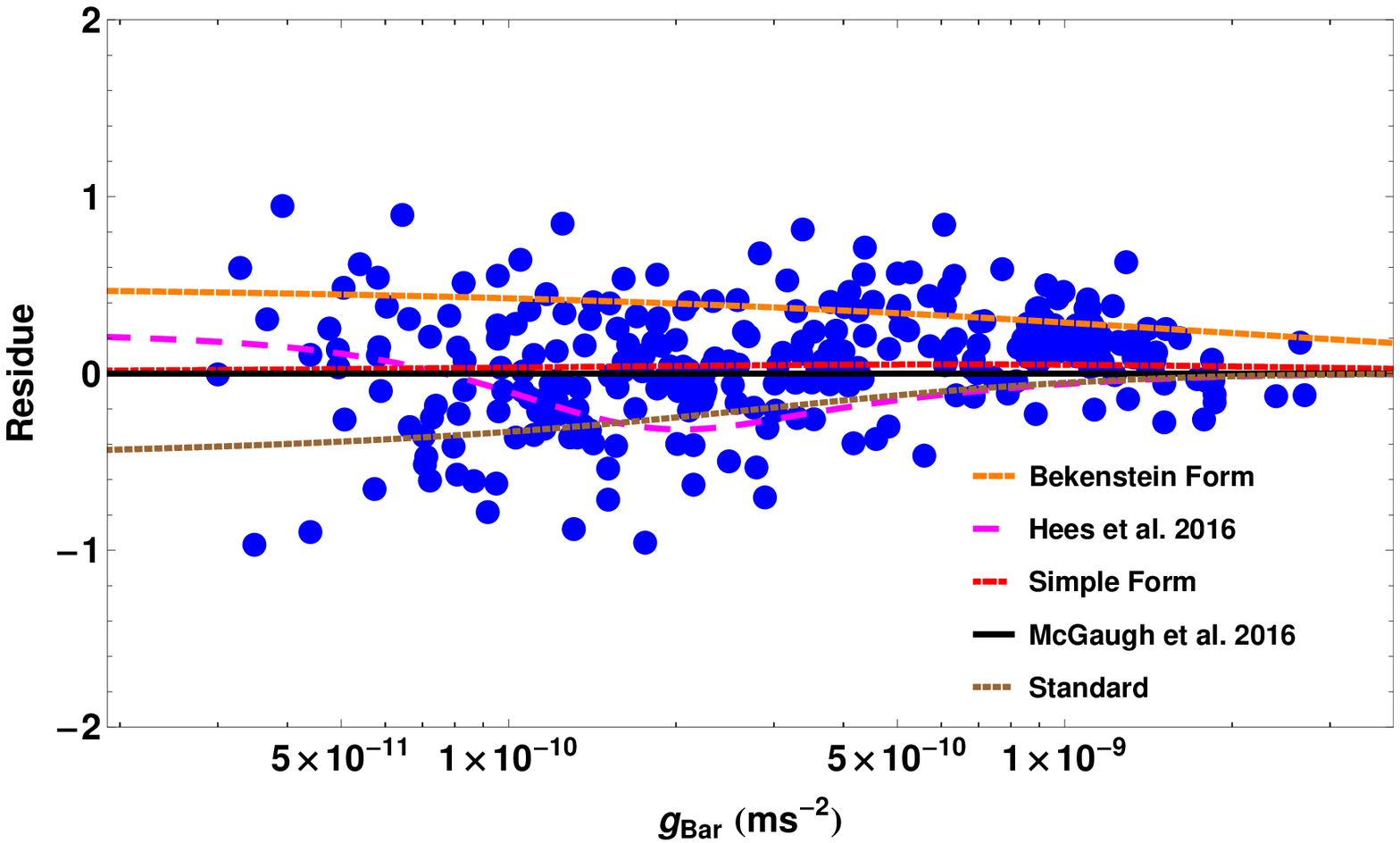}
            \includegraphics[width=\columnwidth]{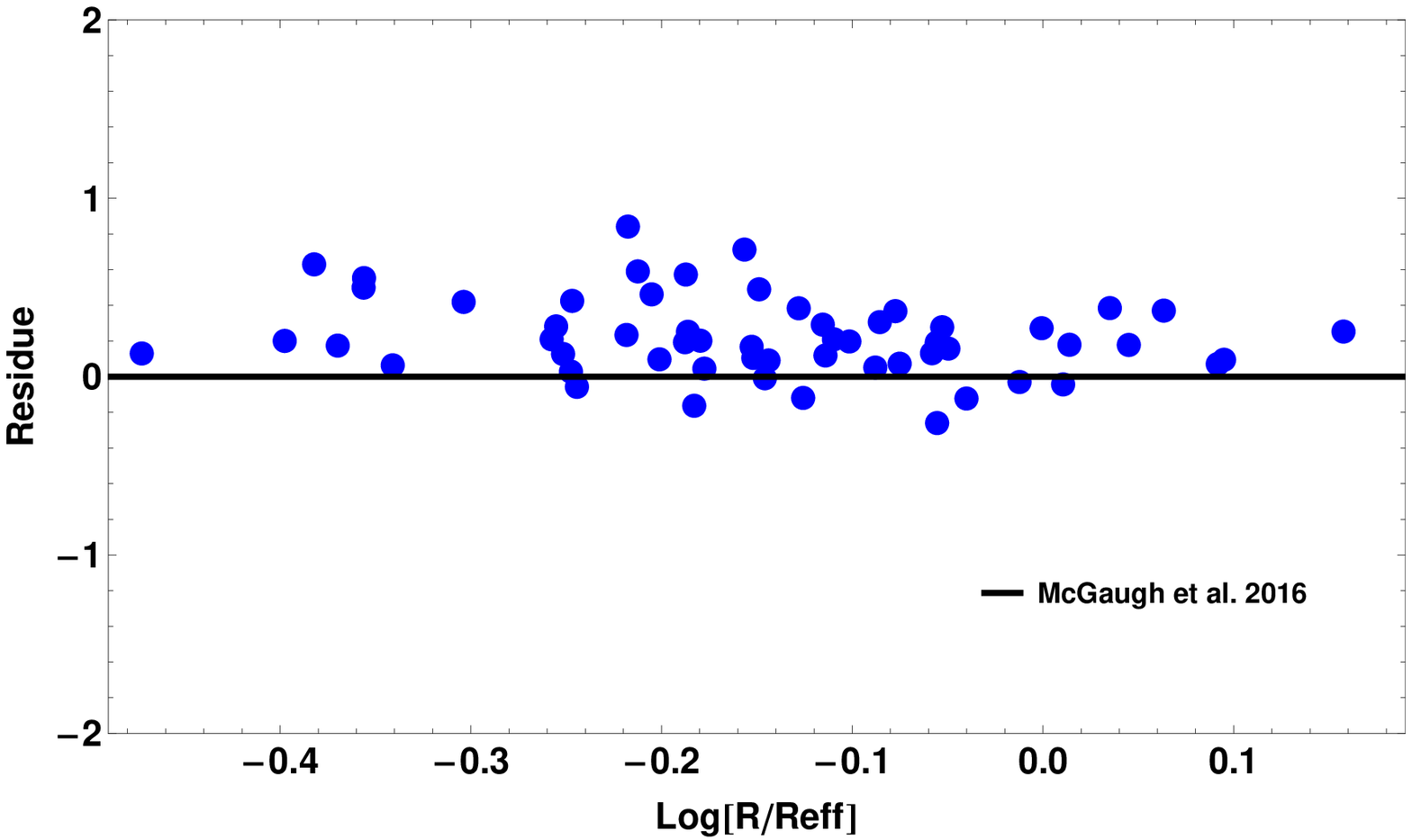}
            \caption{
            Residuals after subtracting MOND interpolating functions from the MDAR.
            Left panel: Blue filled circles are the 285 data points of Einstein rings in this work.
            The horizontal axis is the Newtonian acceleration $g_{\rm Bar}$ (in logarithmic scale) estimated from the baryonic mass $M_{\rm Bar}$.
            Right panel: Blue filled circles are 57 Einstein rings.
            The horizontal axis is the ring radius in terms of effective radius $\log R/R_{\rm eff}$ (in logarithmic scale).
            From left panel to right panel, the interpolating functions are the Bekenstein form, Hees form, simple form, McGaugh form, and standard forms in MOND, respectively.}
            \label{fig:Residues}
        \end{figure*}

    Fig. \ref{fig:Residues} listed the residuals after subtracting the interpolating functions from the MDAR with 285 data points of Einstein rings.
    As expected, simple form and McGaugh form are better than others.
    And Bekenstein form and Standard form seem to be the upper limit and lower limit of MOND prediction.

        \subsection{Surface Mass Density Discrepancy in Einstein Rings}

        \begin{figure}
            \includegraphics[width=\columnwidth]{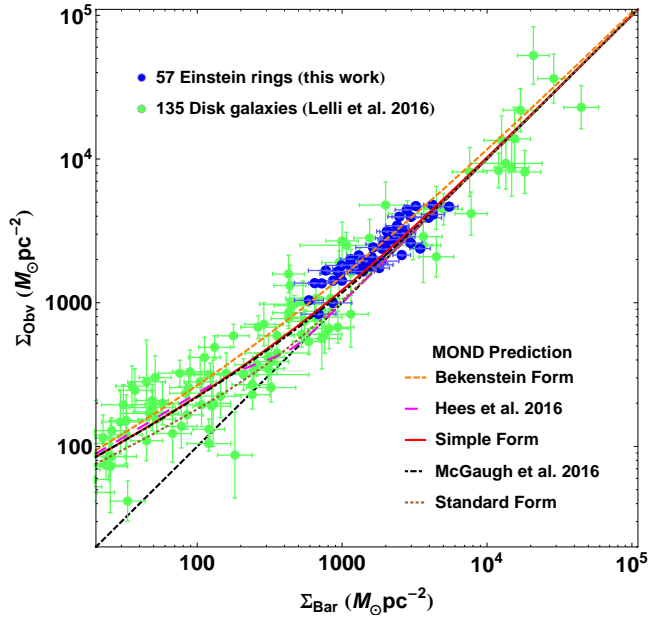}
            \caption{
            Surface mass density discrepancy.
            Blue filled circles are the 57 Einstein rings studied in this work, and green filled circles are the 135 disk galaxies in Lelli et al. (2016).
            The orange dashed, solid, and dotted lines represent the Bekenstein form, simple form, and standard forms in MOND, respectively.
            }\label{fig:SurDis}
        \end{figure}

    Recently, surface mass density discrepancy in disk galaxies is reported in \cite{Lelli16,Milgrom16}.
    The deviation of the surface mass density estimated by dynamics from that by baryons becomes larger as the surface mass density becomes smaller.
    Here, we report a similar discrepancy in Einstein rings (see Fig. \ref{fig:SurDis}).
    Lensing surface mass density $\Sigma_{\rm Obv}$ at the effective radius is obtained by $m_{\rm b}(r)+m_{\rm dm}(r)$ defined earlier.
    Stellar surface mass density $\Sigma_{\rm Bar}$ comes from population synthesis with Salpeter IMF \citep{Auger09}.
    In Fig. \ref{fig:SurDis}, we plot both results from lensing and spiral galaxies \citep{Lelli16} for comparison.
    The two results are consistent.
    The lensing surface mass density in our samples is about $10^{3}$ to $10^{4}$\,M$_{\odot}$\,pc$^{-2}$ which is higher than $\Sigma_0=\mathfrak{a}_0/\pi\,G=276$\,M$_{\odot}$\,pc$^{-2}$.
    Thus, MOND can naturally explain this small discrepancy because these lenses belong to high surface mass density galaxies.
    Although the galaxies belong to this category, the discrepancy trend is still readily observable.
    This is the first time surface mass density discrepancy is discovered in strong gravitational lensing, a relativistic phenomenon.

        \subsection{Relativistic MOND in Gravitational Lensing}

    To consolidate the result of both non-relativistic and relativistic MOND, we compare the lensing mass and dynamical mass in 57 Einstein rings.
    Since SDSS provides the aperture velocity dispersion, the dynamical mass of elliptical galaxies can be computed by the Jeans equation (e.g., \cite{Binney08}, see the appendix also).
    In MOND, both velocity dispersion and gravitational lensing are produced by the same mass distribution (Hernquist model) and the same interpolating function (simple form, Eq. \ref{eq:Simple}).
    As the Hernquist length scale can be estimated by the measured effective radius, the only parameter left is the total mass.

\begin{figure}
        \centering
        \includegraphics[width=\columnwidth]{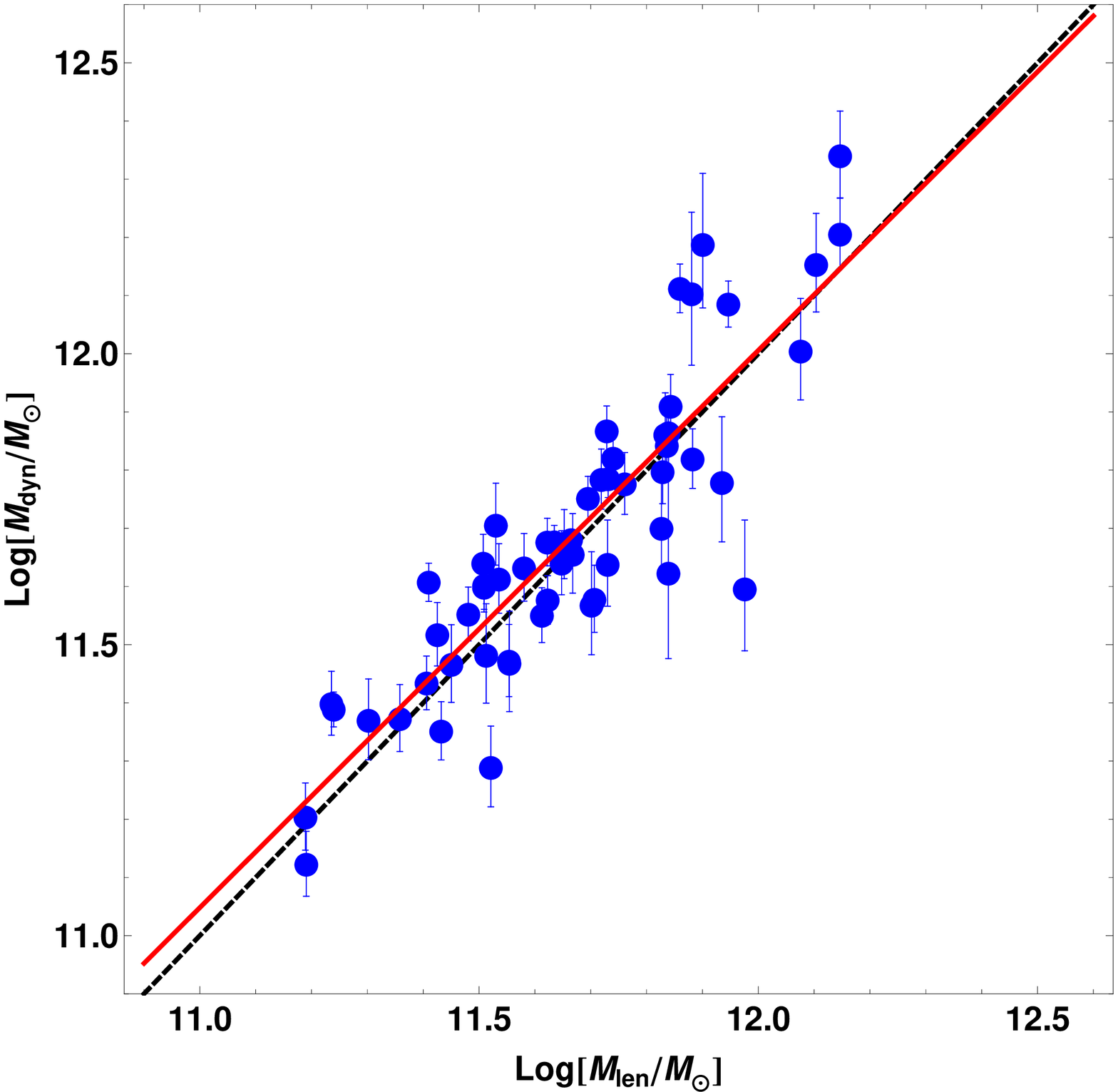}
        \includegraphics[width=\columnwidth]{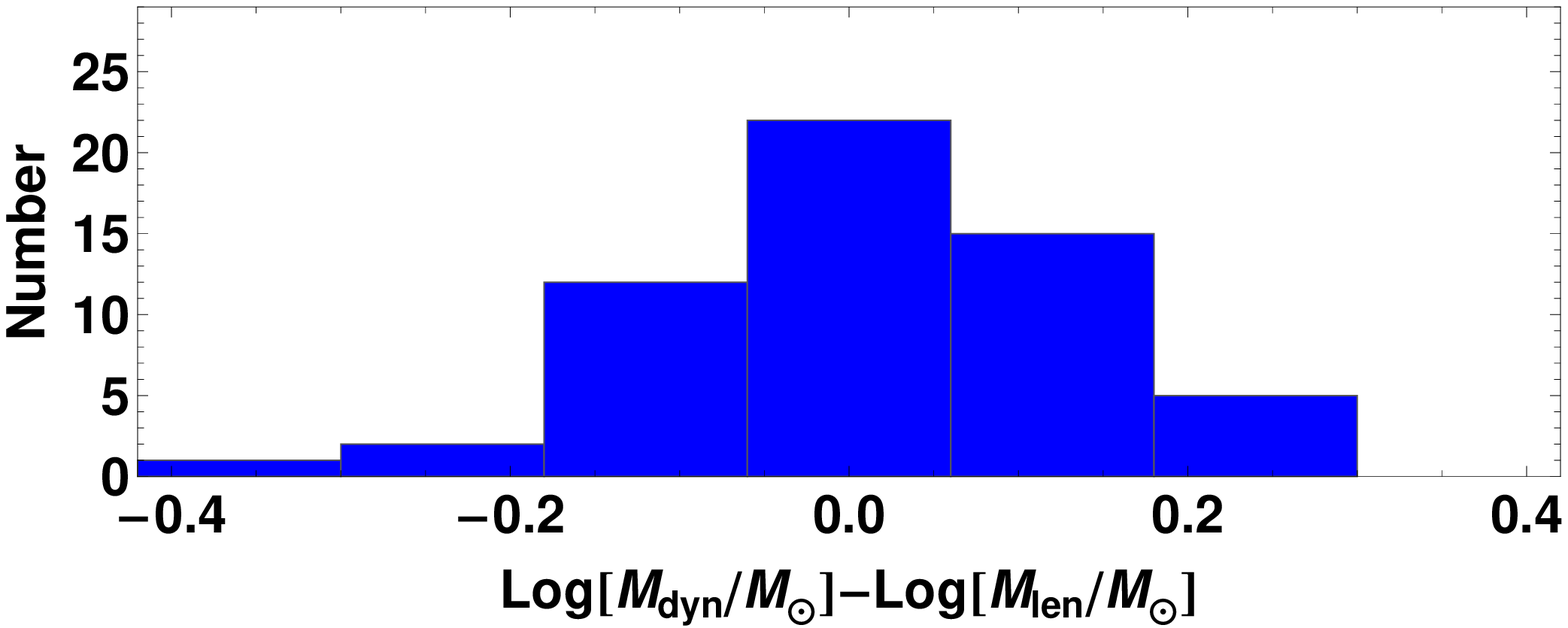}
        \caption{The lensing mass and the dynamical mass of the elliptical galaxies that produce the 57 SLACS Einstein rings in this work.
        The masses are calculated by MOND with simple form.
        Upper panel:
        Red solid line is the best-fit linear correlation: $\log[M_{\rm dyn}/M_{\odot}]=0.96\log[M_{\rm len}/M_{\odot}]+0.51$.
        The black dashed-line denotes the two masses are equal: $M_{\rm dyn}=M_{\rm len}$.
        Error bar in dynamical mass comes from velocity dispersion.
        Lower panel:
        Histogram of the difference between $\log[M_{\rm dyn}/M_{\odot}]$ and $\log[M_{\rm len}/M_{\odot}]$.}
        \label{fig:Mass}
    \end{figure}

    In Fig. \ref{fig:Mass} (upper panel), we compare the total mass calculated from non-relativistic MOND (dynamical mass, $M_{\rm dyn}$) with isotropic velocity distribution and from relativistic MOND (lensing mass, $M_{\rm len}$) of the 57 lensing galaxies in our samples.
    The correlation between these two mass is tight: $\log[M_{\rm dyn}/M_{\odot}]=0.96\log[M_{\rm len}/M_{\odot}]+0.51$.
    The difference between the logarithm of the dynamical mass and the lensing mass is Gaussian (see lower panel of Fig. \ref{fig:Mass}).
    \cite{Sanders14} also gave similar result, but compared mass within Einstein rings and stellar mass instead.

    The correlation between the dynamical mass and lensing mass is still tight if we adopt other interpolating functions, such as Bekenstein form ($\log[M_{\rm dyn}/M_{\odot}]=0.96\log[M_{\rm len}/M_{\odot}]+0.54$) and standard form ($\log[M_{\rm dyn}/M_{\odot}]=0.95\log[M_{\rm len}/M_{\odot}]+0.57$).
    However, different interpolating functions indeed give small differences in mass estimation because the nominal acceleration of our samples is around $10\mathfrak{a}_0$ which is the regime sensitive to interpolating function (see Table \ref{tab:data}).
    If we change the mass model to Jaffe model \citep{Jaffe83}, the lensing mass on average becomes slightly smaller ($5.6\%$ smaller in GR and $5.3\%$ smaller in MOND with simple form).
    The difference in lensing mass of different mass models is less than that of different interpolating functions.
    The dynamical mass calculated from the anisotropic model (Eq. \ref{eq:anisotropic}) is about 3\% to 7\% more when compare with that from the isotropic model.
    Moreover, the lensing mass and dynamical mass calculated under Hernquist model and simple form agree well with the stellar mass with Salpeter IMF (see Table \ref{tab:data}).
    Finally, the mass-to-light ratio of lensing mass in Relativistic MOND in simple form and V-band luminosity ranges from 2.2 to 7.6 with the average around 4.4 in units of ${\rm M}_{\odot}/{\rm L}_{\odot}$ which is the same as the result of \cite{Sanders14}.

    When the surface mass density $\Sigma$ is estimated by the stellar mass at the effective radius, $\Sigma/\Sigma_0>1$ for all our sample galaxies, and the average is $\langle\Sigma/\Sigma_0\rangle=7.1$.
    Thus, our samples belong to the high surface mass density category.
    From our analysis, the lensing mass of relativistic MOND in simple form is smaller than that from GR by about $23\%\pm5\%$, i.e., the mass discrepancy is small, as expected.
    The acceleration in relativistic MOND at the effective radius is also larger than $\mathfrak{a}_0$ with an average
    $\langle a/\mathfrak{a}_0\rangle=7.3$, which is consistent with the surface mass density estimation, see Table \ref{tab:data}.

    Our analysis on 57 Einstein rings shows the existence of the MDAR and surface mass density discrepancy in gravitational lensing (a relativistic phenomenon).
    MOND can provide a way to understand the MDAR and surface mass density discrepancy.
    We also show the consistency between relativistic MOND and non-relativistic MOND.

    \begin{table*}
    \sisetup{table-number-alignment = center}
    \centering
    \tabcolsep=4pt
    \setlength{\extrarowheight}{3pt}
    \caption[\textbf{The sample of 57 Einstein rings by elliptical lenses.}]{\textbf{The sample of 57 Einstein rings by elliptical lenses.}}\label{tab:data}
        \begin{tabular}{
        c
        S[table-figures-integer=1,table-figures-decimal=2]
        S[table-figures-integer=2,table-figures-decimal=2]
        S[table-figures-decimal=0,separate-uncertainty,table-figures-uncertainty=1]
        S[separate-uncertainty,table-figures-uncertainty=1]
        S[table-figures-integer=2,table-figures-decimal=2]
        c
        S[table-figures-decimal=1,separate-uncertainty,table-figures-uncertainty=1]
        S[table-figures-decimal=1,separate-uncertainty,table-figures-uncertainty=1]
        S[table-figures-decimal=1,separate-uncertainty,table-figures-uncertainty=1]
        S[table-figures-integer=1,table-figures-decimal=1]
        }
        \hline
        {Name} & {$R_{\rm Obv}$} & {$R_{\rm eff}$} & {$\sigma$} & {$M_{\rm Bar}$} & {$M_{\rm len}$} & {$M_{\rm dyn}$} &	 {$\frac{\theta_{\rm Obv}}{\theta_{\rm Bar}}$} & {$\frac{M_{\rm Obv}}{M_{\rm Bar}}$} & {$\frac{g_{\rm N}}{\mathfrak{a}_0}$} & {$\frac{a}{\mathfrak{a}_0}$} \\
         & & &	& {\footnotesize{IMF}} & {\footnotesize{MOND}} & {\footnotesize{MOND}} & & {\footnotesize{$\frac{\rm ISO}{\rm IMF}$}} & {\footnotesize{IMF}} & {\footnotesize{IMF}} \\
         &	{kpc} &{kpc} & {\si{\kilo\metre\per\second}} & {\footnotesize{$\log{M}_{\odot}$}} & {\footnotesize{$\log{M}_{\odot}$}} & {\footnotesize{$\log{M}_{\odot}$}} & & & & \\
        {(1)} & {(2)} & {(3)} & {(4)} &{(5)} & {(6)} & {(7)} & {(8)} & {(9)} & {(10)} & {(11)}  \\						  				
        \hline																																																																																 J0008$-$0004& 6.59 & 9.45 & 193(36) & 11.64(14) & 11.84 & 11.62$^{+0.17}_{-0.15}$ & 2.1(7) & 2.0(6) & 2.3(8) & 3.7 \\
        J0029$-$0055& 3.48 & 7.63 & 229(18) & 11.58(13) & 11.53 & 11.70$^{+0.07}_{-0.07}$ & 1.1(3) & 1.1(3) & 3.1(9) & 2.8 \\
        J0037$-$0942& 4.95 & 5.66 & 279(10) & 11.73(06) & 11.73 & 11.78$^{+0.03}_{-0.03}$ & 1.2(2) & 1.2(2) & 8.0(11) & 8.0 \\
        J0044$+$0113& 1.72 & 4.03 & 266(13) & 11.47(09) & 11.51 & 11.60$^{+0.04}_{-0.04}$ & 1.2(2) & 1.2(2) & 8.7(18) & 9.5 \\
        J0157$-$0056& 4.89 & 11.1 & 295(47) & 11.74(10) & 11.88 & 12.10$^{+0.14}_{-0.12}$ & 1.7(4) & 1.7(4) & 2.1(5) & 3.0 \\
        J0216$-$0813&	5.53 & 11.13 & 333(23)&	12.03(07) &	12.15 & 12.20$^{+0.06}_{-0.06}$ & 1.5(2) & 1.5(2) & 4.1(7) & 5.4 \\
        J0252$+$0039&	4.4	 & 5.74	& 164(12) &	11.46(13) &	11.52 & 11.29$^{+0.07}_{-0.07}$ & 1.5(4) & 1.4(4) & 4.2(13) & 4.8 \\
        J0330$-$0020&	5.45 & 4.38	& 212(21) &	11.58(09) &	11.55 & 11.47$^{+0.09}_{-0.08}$ & 1.2(3) & 1.2(2) & 9.5(20) & 8.9 \\
        J0728$+$3835&	4.21 & 5.89	& 214(11) &	11.69(12) & 11.61 & 11.55$^{+0.05}_{-0.05}$ & 1.0(3) & 1.0(3) & 6.8(19) & 5.7 \\
        J0737$+$3216&	4.66 & 8.18	& 338(16) &	11.96(07) &	11.86 & 12.11$^{+0.04}_{-0.04}$ & 0.9(2) & 1.0(2) & 6.5(11) & 5.2 \\
        J0819$+$4534&	2.73 & 6.2	& 225(15) &	11.40(08) & 11.54 & 11.61$^{+0.06}_{-0.06}$ & 1.6(3) & 1.6(3) & 3.1(6) & 4.3 \\
        J0822$+$2652&	4.45 & 6.73	& 259(15) &	11.69(13) &	11.72 & 11.78$^{+0.05}_{-0.05}$ & 1.3(4) & 1.3(4) & 5.2(16) & 5.5 \\
        J0903$+$4116&	7.23 & 9.71	& 223(27) &	11.84(14) &	11.93 & 11.78$^{+0.11}_{-0.10}$ & 1.6(3) & 1.5(3) & 3.1(6) & 4.4 \\
        J0935$-$0003&	4.26 &10.27	& 396(35) &	11.96(07) &	12.15 & 12.34$^{+0.08}_{-0.07}$ & 1.7(3) & 1.7(3) & 4.1(7) & 6.4 \\
        J0936$+$0913&	3.45 & 6.1	& 243(11) &	11.68(12) &	11.62 & 11.68$^{+0.04}_{-0.04}$ & 1.0(3) & 1.1(3) & 6.2(17)	 & 5.4 \\
        J0946$+$1006&	4.95 & 8.17	& 263(21) &	11.59(12) & 11.83 & 11.86$^{+0.07}_{-0.07}$ & 2.1(6) & 2.0(6) & 2.8(8) & 4.9 \\
        J0956$+$5100&	5.05 & 8.1	& 334(15) &	11.81(08) &	11.95 & 12.08$^{+0.04}_{-0.04}$ & 1.6(3) & 1.6(3) & 4.7(9) & 6.5 \\
        J0959$+$0410&	2.24 & 2.83	& 197(13) &	11.15(06) &	11.19 & 11.20$^{+0.06}_{-0.06}$ & 1.3(2) & 1.2(2) & 8.4(12) & 9.2 \\
        J1016$+$3859&	3.13 & 4.07	& 247(13) &	11.48(12) &	11.48 & 11.55$^{+0.05}_{-0.05}$ & 1.2(3) & 1.2(3) & 8.7(24) & 8.7 \\
        J1020$+$1122&	5.12 & 6.59	& 282(18) &	11.80(12) &	11.84 & 11.86$^{+0.06}_{-0.05}$ & 1.3(4) & 1.3(4) & 7.0(19) & 7.6 \\
        J1023$+$4230&	4.5	 & 5.48	& 242(15) &	11.57(12) &	11.65 & 11.64$^{+0.06}_{-0.05}$ & 1.5(4) & 1.4(4) & 5.9(16) & 7.1 \\
        J1100$+$5329&	7.02 & 9.89	& 187(23) &	11.84(07) &	11.98 & 11.59$^{+0.12}_{-0.11}$ & 1.7(3) & 1.7(3) & 3.4(5) & 4.6 \\
        J1106$+$5228&	2.17 & 2.38	& 262(9)  &	11.37(06) &	11.24 & 11.39$^{+0.03}_{-0.03}$ & 0.8(1) & 0.9(1) & 19.8(27) & 14.7 \\
        J1112$+$0826&	6.19 & 5.35	& 320(20) &	11.73(08) &	11.84 & 11.91$^{+0.06}_{-0.05}$ & 1.6(3) & 1.4(3) & 9.0(17) & 11.7 \\
        J1134$+$6027&	2.93 & 5.23	& 239(11) &	11.51(12) &	11.51 & 11.60$^{+0.04}_{-0.04}$ & 1.2(3) & 1.2(3) & 5.7(16) & 5.7 \\
        J1142$+$1001&	3.52 & 4.31	& 221(22) &	11.55(08) &	11.51 & 11.48$^{+0.09}_{-0.08}$ & 1.1(2) & 1.1(2) & 9.1(17) & 8.4 \\
        J1143$-$0144&	3.27 & 5.02	& 269(5)  &	11.60(09) &	11.66 & 11.68$^{+0.02}_{-0.02}$ & 1.3(3) & 1.3(3) & 7.6(16) & 8.8 \\
        J1153$+$4612&	3.18 & 3.08	& 226(15) &	11.33(13) &	11.36 & 11.37$^{+0.06}_{-0.06}$ & 1.2(4) & 1.2(4) & 10.8(32) & 11.5 \\
        J1204$+$0358&	3.68 & 2.98	& 267(17) &	11.45(06) &	11.43 & 11.52$^{+0.06}_{-0.05}$ & 1.1(2) & 1.1(2) & 15.2(21) & 14.4 \\
        J1205$+$4910&	4.27 & 6.07	& 281(13) &	11.72(06) &	11.74 & 11.82$^{+0.04}_{-0.04}$ & 1.2(2) & 1.2(2) & 6.8(9) & 7.1 \\
        J1213$+$6708&	3.13 & 3.22	& 292(11) &	11.49(09) &	11.41 & 11.61$^{+0.03}_{-0.03}$ & 1.0(2) & 1.0(2) & 14.3(30) & 11.9 \\
        J1218$+$0830&	3.47 & 6.28	& 219(10) &	11.59(08) &	11.62 & 11.58$^{+0.04}_{-0.04}$ & 1.3(2) & 1.3(2) & 4.7(9) & 5.1 \\
        J1250$+$0523&	4.18 & 4.75	& 252(14) &	11.77(07) &	11.51 & 11.64$^{+0.05}_{-0.05}$ & 0.7(1) & 0.8(2) & 12.5(20) & 6.8 \\
        J1306$+$0600&	3.87 & 3.57	& 237(17) &	11.43(08) &	11.55 & 11.47$^{+0.06}_{-0.06}$ & 1.6(3) & 1.4(3) & 10.1(19) & 13.4 \\
        J1313$+$4615&	4.25 & 4.8	& 263(18) &	11.58(08) &	11.65 & 11.67$^{+0.06}_{-0.06}$ & 1.4(3) & 1.3(2) & 7.9(15) & 9.3 \\
        J1318$-$0313&	6.01 & 9.25	& 213(18) &	11.67(09) & 11.83 & 11.70$^{+0.08}_{-0.07}$ & 1.8(4) & 1.8(4) & 2.6(5) & 3.8 \\
        J1402$+$6321&	4.53 & 7.49	& 267(17) &	11.79(06) &	11.84 &	 11.84$^{+0.06}_{-0.05}$ & 1.3(2) & 1.3(2) & 5.3(7) & 5.9 \\
        J1403$+$0006&	2.62 & 3.5	& 213(17) &	11.44(08) &	11.30 & 11.37$^{+0.07}_{-0.07}$ & 0.9(2) & 0.9(2) & 10.1(19) & 7.8 \\
        J1416$+$5136&	6.08 & 4.23	& 240(25) &	11.64(08) & 11.70 &	 11.57$^{+0.09}_{-0.08}$ & 1.4(3) & 1.3(3) & 11.7(22) & 13.5 \\
        J1430$+$4105&	6.53 &10.65	& 322(32) &	11.93(11) &	12.10 & 12.15$^{+0.09}_{-0.08}$ & 1.8(5) & 1.7(4) & 3.6(9) & 5.4 \\
        J1436$-$0000&	4.8	 & 6.81	& 224(17) &	11.69(09) &	11.67 & 11.65$^{+0.07}_{-0.07}$ & 1.2(2) & 1.2(2) & 5.1(10) & 4.8 \\
        J1451$-$0239&	2.33 & 3.55	& 223(14) &	11.39(06) &	11.24 & 11.40$^{+0.06}_{-0.05}$ & 0.9(1) & 0.8(1) & 9.3(13) & 6.5 \\
        \hline
    \end{tabular}
    \end{table*}

    \begin{table*}
        \contcaption{}
    \sisetup{table-number-alignment = center}
    \centering
    \tabcolsep=4pt
    \setlength{\extrarowheight}{3pt}
        \begin{tabular}{
        c
        S[table-figures-integer=1,table-figures-decimal=2]
        S[table-figures-integer=2,table-figures-decimal=2]
        S[table-figures-decimal=0,separate-uncertainty,table-figures-uncertainty=1]
        S[separate-uncertainty,table-figures-uncertainty=1]
        S[table-figures-integer=2,table-figures-decimal=2]
        c
        S[table-figures-decimal=1,separate-uncertainty,table-figures-uncertainty=1]
        S[table-figures-decimal=1,separate-uncertainty,table-figures-uncertainty=1]
        S[table-figures-decimal=1,separate-uncertainty,table-figures-uncertainty=1]
        S[table-figures-integer=1,table-figures-decimal=1]
        }
        \hline
        {Name} & {$R_{\rm Obv}$} & {$R_{\rm eff}$} & {$\sigma$} & {$M_{\rm Bar}$} & {$M_{\rm len}$} & {$M_{\rm dyn}$} &	 {$\frac{\theta_{\rm Obv}}{\theta_{\rm Bar}}$} & {$\frac{M_{\rm Obv}}{M_{\rm Bar}}$} & {$\frac{g_{\rm N}}{\mathfrak{a}_0}$} & {$\frac{a}{\mathfrak{a}_0}$} \\
         & & &	& {\footnotesize{IMF}} & {\footnotesize{MOND}} & {\footnotesize{MOND}} & & {\footnotesize{$\frac{\rm ISO}{\rm IMF}$}} & {\footnotesize{IMF}} & {\footnotesize{IMF}} \\
         &	{kpc} &{kpc} & {\si{\kilo\metre\per\second}} & {\footnotesize{$\log{M}_{\odot}$}} & {\footnotesize{$\log{M}_{\odot}$}} & {\footnotesize{$\log{M}_{\odot}$}} & & & & \\
        {(1)} & {(2)} & {(3)} & {(4)} &{(5)} & {(6)} & {(7)} & {(8)} & {(9)} & {(10)} & {(11)}  \\						  				
        \hline
        J1525$+$3327&	6.55 &11.79	& 264(26) &	12.02(09) &	12.08 & 12.00$^{+0.09}_{-0.08}$ & 1.4(3) & 1.4(3) & 3.6(7) & 4.1 \\
        J1531$-$0105&	4.71 & 5.28	& 279(12) &	11.68(09) &	11.70 & 11.75$^{+0.04}_{-0.04}$ & 1.2(3) & 1.2(3) & 8.2(17) & 8.5 \\
        J1538$+$5817&	2.5	 & 2.44	& 189(12) &	11.28(08) &	11.19 & 11.12$^{+0.06}_{-0.05}$ & 1.0(2) & 1.0(2) & 15.3(28) & 12.5 \\
        J1614$+$4522&	2.54 & 7.54	& 182(13) &	11.47(12) &	11.45 & 11.47$^{+0.07}_{-0.06}$ & 1.2(3) & 1.2(3) & 2.5(7) & 2.4 \\
        J1621$+$3931&	4.97 & 5.65	& 236(20) &	11.70(07) &	11.73 & 11.64$^{+0.08}_{-0.07}$ & 1.3(2) & 1.3(2) & 7.5(12) & 8.1 \\
        J1627$-$0053&	4.18 & 6.44	& 290(14) &	11.70(09) &	11.73 & 11.87$^{+0.04}_{-0.04}$ & 1.3(3) & 1.3(3) & 5.8(12) & 6.2 \\
        J1630$+$4520&	6.91 & 6.23	& 276(16) &	11.86(07) &	11.88 & 11.82$^{+0.05}_{-0.05}$ & 1.3(2) & 1.2(2) & 8.9(14) & 9.4 \\
        J1636$+$4707&	3.96 & 5.96	& 231(15) &	11.63(08) &	11.58 & 11.63$^{+0.06}_{-0.06}$ & 1.1(2) & 1.1(2) & 5.8(11) & 5.1 \\
        J1644$+$2625&	3.07 & 3.65	& 229(12) &	11.43(08) &	11.41 & 11.43$^{+0.05}_{-0.04}$ & 1.1(2) & 1.1(2) & 9.7(18) & 9.2 \\
        J2238$-$0754&	3.08 & 4.29	& 198(11) &	11.45(06) &	11.43 & 11.35$^{+0.05}_{-0.05}$ & 1.1(2) & 1.1(2) & 7.3(10) & 7.0 \\
        J2300$+$0022&	4.51 & 5.39	& 279(17) &	11.65(07) &	11.76 & 11.78$^{+0.05}_{-0.05}$ & 1.5(2) & 1.4(2) & 7.4(12) & 9.5 \\
        J2303$+$1422&	4.35 & 7.68	& 255(16) &	11.71(06) &	11.83 & 11.80$^{+0.06}_{-0.05}$ & 1.6(2) & 1.5(2) & 4.2(6) & 5.5 \\
        J2321$-$0939&	2.47 & 6.17	& 249(8)  &	11.60(08) &	11.63 & 11.68$^{+0.03}_{-0.03}$ & 1.2(2) & 1.2(2) & 5.0(9) & 5.4 \\
        J2341$+$0000&	4.5	 & 7.15	& 207(13) &	11.73(08) &	11.71 & 11.58$^{+0.06}_{-0.06}$ & 1.2(2) & 1.2(2) & 5.0(9) & 4.8 \\
        J2347$-$0005&	6.1	 & 6.11	& 404(59) &	11.83(08) &	11.90 & 12.19$^{+0.12}_{-0.11}$ & 1.4(3) & 1.3(2) & 8.7(16) & 10.2 \\
    \hline
    \end{tabular}
    \begin{tablenotes}
    \item[]
            (1) Name of galaxy,                                                         \\
            (2) the radius of Einstein ring in kpc from SLACS \citep{Auger09},           \\
            (3) I-band effective radius in kpc from SLACS \citep{Auger09},               \\
            (4) velocity dispersion averaged over aperture radius from SLACS \citep{Auger09},   \\
            (5) mass estimated from population synthesis models with Salpeter IMF \citep{Auger09},   \\
            (6) fitting mass of the lens in relativistic MOND in simple form from gravitational lensing,    \\
            (7) fitting mass of the galaxy in MOND in simple form from dynamics,                            \\
            (8) acceleration discrepancy between Einstein radius from the observation and Einstein radius from stellar mass with Salpeter IMF,    \\
            (9) mass discrepancy (acceleration discrepancy) between lensing mass in GR with singular isothermal (ISO) model and stellar mass with Salpeter IMF,    \\
            (10) Newtonian acceleration estimated by the Salpeter IMF at effective radius in unit of MOND acceleration constant $\mathfrak{a}_0=1.2\times10^{-10}$ ms$^{-2}$,  \\
            (11) the acceleration estimated by lensing mass in relativistic MOND in simple form at the effective radius. \\
            All masses are in unit of $\log{M}_{\odot}$.
            Hernquist profile \citep{Hernquist90} is adopted for luminous matter distribution.
    \end{tablenotes}
    \end{table*}

    \appendix
    \section{Aperture Velocity Dispersion}\label{sec:Jeans}

    For simplicity, we model an elliptical galaxy as a spherically symmetric stellar system.

        The velocity dispersion of a spherically symmetric stellar system in equilibrium is governed by the Jeans equation in spherical coordinates \citep{Binney08},
        \begin{equation}
        \label{eq:Jeans}
        \frac{{\rm d}(\rho\sigma_r^2)}{{\rm d}r}+\frac{2\beta_a}{r}\rho\sigma_r^2=-\rho g\,,
        \end{equation}
        where $\beta_a=1-(\sigma_t^2/\sigma_r^2)$ is the anisotropy parameter\,($\beta_a=0$ for the isotropy case).

        The velocity dispersion averaged over radius measured along the line of sight at projected aperture radius $R$ is given by
        \begin{equation}
        \label{eq:vel_I}
        \sigma_S^2(R)=\frac{4\pi}{S(R)}\int_{0}^{R}\int_{R'}^{\infty}\sigma_r^2\left(1-\beta_a(r)\frac{R'^2}{r^2}\right)
        \frac{\rho(r)rR'}{\sqrt{r^2-R'^2}}\,{\rm d}r{\rm d}R'\,,
        \end{equation}
        where the cumulative surface density is
        \begin{equation}
        \label{eq:surface}
        S(R)=4\pi\int_{R}^{\infty}\rho(r)r^2{\rm d}r-4\pi\int_{R}^{\infty}\rho(r) r\sqrt{r^2-R^2}\,{\rm d}r\,.
        \end{equation}

        In this paper, beside isotropic model (i.e., $\beta_a=0$) we also consider a particular anisotropic model

        \begin{equation}
        \label{eq:anisotropic}
        \beta_a(r)=\frac{r^2}{r_{\rm a}^2+r^2}\,.
        \end{equation}
        This anisotropic model can be formed by dissipationless collapse systems when $r_{\rm a}$ equals to three times the effective radius $R_{\rm eff}$ ($r_{\rm a}=3R_{\rm eff}$) \citep{vanAlbada82,MS03}.

\section*{ACKNOWLEDGEMENTS}
    We are grateful to the anonymous reviewer for comments and suggestions to improve our work.
    This work is supported in part by the Taiwan Ministry of Science and Technology grants MOST 104-2923-M-008-001-MY3 and MOST 105-2112-M-008-011-MY3.

\end{document}